# Impact of fluctuations in social contacts on the spread of epidemics

S. P. Fisenko and P. S. Grinchuk
A.V. Luikov Heat and Mass Transfer Institute of National Academy of Sciences
15, P. Brovka St., 220072, Minsk, Belarus. fsp@hmti.ac.by

*The mathematical model of the spread of the epidemic is developed, which considers fluctuations in the number of human social contacts. The model is based on the logistic equation. The oscillation in the number of social contacts within one week is considered as the main source of fluctuations. The time-fluctuating variable of the number of contacts is represented as the sum of the first two terms of the Fourier series expansion of number of contacts. It is shown that the largest fluctuation amplitudes occur near the maximum rate of the epidemic spread. After that, the effect of fluctuations decreases until the end of the epidemic.*

**Introduction**. One of the features of the intense spread of the COVID-19 epidemic around the world is the large fluctuations in daily cases. Currently available primary statistical data on the spread of the COVID-19 pandemic [1] and simulation results show that the classical mathematical model of the spread of the epidemic in a single country, based on a nonlinear ordinary differential equation [2, 3] has a fundamental drawback. This drawback of the classical mathematical model is that it does not describe an important effect, namely, fluctuations (oscillations) in the number of sick people every day. The biggest reason for that seems to be the appearance of the fluctuation mode for the number of human social contacts, especially on weekends, that are related to the organization of the life of a civilized human society within a period of 7 days. As a result of this phenomenon fluctuations in the number of sick people are observed.

There are many papers that consider multivariate complex models of the spread of epidemics with a large number of empirical parameters. In such works the influence of fluctuations of various nature (initial conditions, random vaccination, etc.) on the spread of epidemic diseases was considered [4-14]. Many of these papers are devoted to the mathematical issues of solution stability and solution behavior in various epidemiological models.

The aim of this paper is to incorporate fluctuation effects related to the social contacts in a simple mathematical model of an epidemic spread in human society.

For a more complete description of the spread of the epidemic, it is necessary to use a more general theoretical description of the dynamic processes underlying the spread of the COVID -19 epidemics. Human society is a much more complex system than ordinary objects of statistical physics [15]. Nevertheless, it is possible to propose the use of the probability density *f*(y, t) of the number of infected people y at time *t*. What kinetic equation governs the evolution of the probability density function of the number of infected people? At present, as far as we know, such an equation has not yet been obtained. An alternative approach is to introduce random Langevin's sources into equation (1).

Next, for the first time we use a simpler approach to the fluctuation of infected peoples based on the logistic equation with an oscillating propagation coefficient. The logistic equation used to describe the spread of an epidemic is

$$\frac{dy}{dt} = kyb\left(1 - \frac{y}{N}\right). \qquad (1)$$

Here $y$ is the number of sick people, $kb$ is the product of the average probability of being infected $k$, and $b$ is the average number of contacts of the patient with other people, $N$ is the population size. For a constant value of the product kb in particular the exact solution of the logistic equation is given in [2]. The $kb$ value is called the epidemic reproduction number and is often denoted as $R_0$.

## Oscillations of reproduction number

Let assume that the reproduction number $R_0$ for a population is not constant value and fluctuate in time around a certain average value. The reason for fluctuations in the reproduction number is fluctuations in the number of human contacts on different days. The exact solution of the logistic equation with variable reproduction number has the form

$$y(t) = \frac{N}{1 + CN\exp\left(-\int_0^t kb(t_1)dt_1\right)}. \quad (2)$$

Note that for the initial condition y(0) = 1 we have the following expression for the number of sick people:

$$y(t) = \frac{N}{1 + (N-1)\exp\left(-\int_0^t kb(t_1)dt_1\right)}. \quad (3)$$

Let us formally expand the dependence of the parameter $b$(t) into the Fourier series over an interval of 7 days. Below we consider only the first three terms of this expansion. Then we have the expression:

$$b(t) = b_a + b_1 \sin(\omega t + \varphi). \quad (4)$$

Here $b_a$ is a constant value, which depends on a whole set of data, φ is a phase variable. The second term on the right side of expression (4) oscillates with the fundamental frequency ω, so that the characteristic time is 1/ω. Obviously, expression (4) makes it possible to consider a more detailed picture of the spread of the epidemic than most theoretical works that simulate the spread of the epidemic. For simplicity of analysis the influence of the phase variable in expression (4) will be neglected. Note that, as follows from our subsequent results, the contributions of higher harmonics 2ω, 3ω are mutually averaged and can be neglected. It is useful to note that, in fact, we are expanding the product *kb* in a Fourier series, although for definiteness we will consider below only fluctuations in the number of contacts.

Now, taking into account (4), we can analytically calculate the integral in (3). For the number of infected people $y$(t) we have the following formula:

$$y(t) = \frac{N}{1 + (N-1)\exp - (kb_a t)\exp\left[kb_1\left(\frac{\cos(\omega t) - 1}{\omega}\right)\right]} \tag{5}$$

Obviously, the expression (5) gives the oscillating character of the number of infected people in relation to the non- oscillating curve $y_s$(t). This curve is determined by the expression (6), which is the non-oscillating part of the Exp. (5):

$$y_s(t) = \frac{N}{1 + (N-1)\exp - (kb_a t)}. \tag{6}$$

For sufficiently long times of the spread of the epidemics from exps. (5) and (6) we obtain an approximate formula for the deviation of the number of cases from the non-oscillating regime $y$(t) - $y_s$(t):

$$y(t) - y_s(t) \approx N^2 \exp(-kb_a t)\left\{\exp\left[kb_1\left(\frac{\cos(\omega t) - 1}{\omega}\right)\right] - 1\right\}. \tag{7}$$

It is important to note that at these times the deviation value y (t) - $y_s$ (t) is directly proportional to $N^2$ and periodically changes its sign as well as decreases with time. Thus, at the end of the epidemic, fluctuation effects decrease. This result contradicts the statement in [16] that fluctuations intensify at the end of an epidemic.

Similarly, for the beginning of the epidemic spread from Exps. (5) and (6) we have an approximate formula

$$y(t) - y_s(t) \approx \exp(kb_a t)\left[\frac{1}{\exp\left[kb_1\left(\frac{\cos(\omega t) - 1}{\omega}\right)\right]} - 1\right]. \tag{8}$$

It can be seen that the magnitude of the deviation from the non-oscillating regime exponentially increases with time, does not depend on the population size, and periodically changes its sign. Thus, expressions (6-8) clarify the statement made in [5] about the manifestation of fluctuations (oscillations) during the spread of an epidemic.

Results of the numerical solution of the logistic equation with an oscillating value of $b$(t) are shown in Fig. 1. The difference (y(t) - $y_s$) is normalized to the population size $N$. Note that for a population with $N = 2 \cdot 10^6$ people, a 3% deviation means 60 thousand sick people.

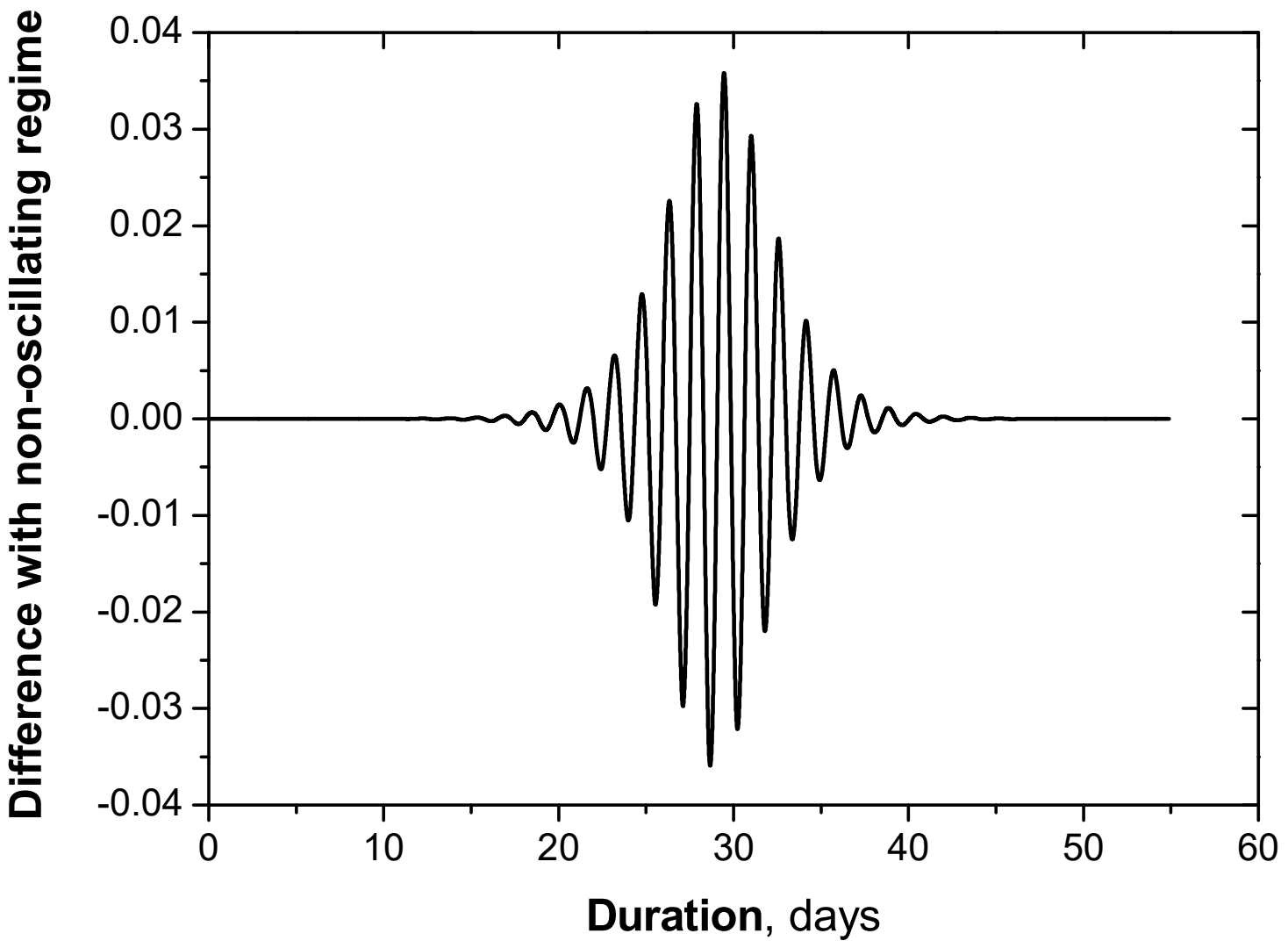

Fig.1 Fluctuations influence on the normalized number of cases over time

$kb_a = 0.5$; $kb_1 = 0.6$; $\omega = 4$.

In principle, all parameters for expression (4) should be determined either from Fourier analysis of statistical data on the number of sick people. However, it is useful to carry out a parametric simulation of our mathematical model. In Fig. 2 the maximum of deviation amplitudes $(y\text{-}y_s)/y_s$ for two values of frequencies $\omega$ and for different ratios of the two parameters $b_1/b_a$ is shown. It can be seen that with an increase in the frequency of fluctuations, the contribution of fluctuations is effectively averaged and the amplitude of fluctuations (oscillations) in the number of sick people decreases.

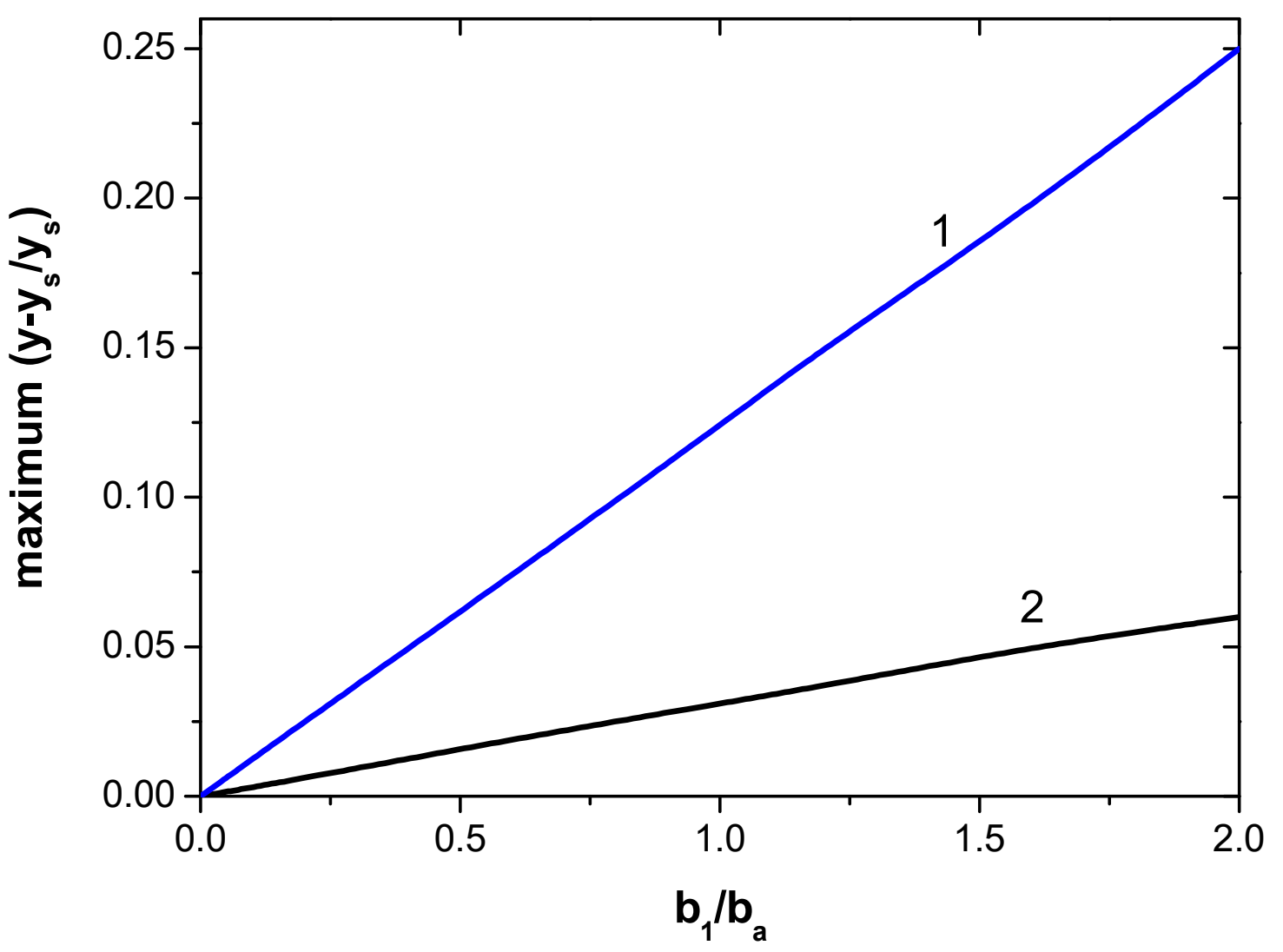

Figure 2. Normalized maximum of deviations from the non-oscillating regime

Curve 1 for ω = 1, curve 2 for ω = 4

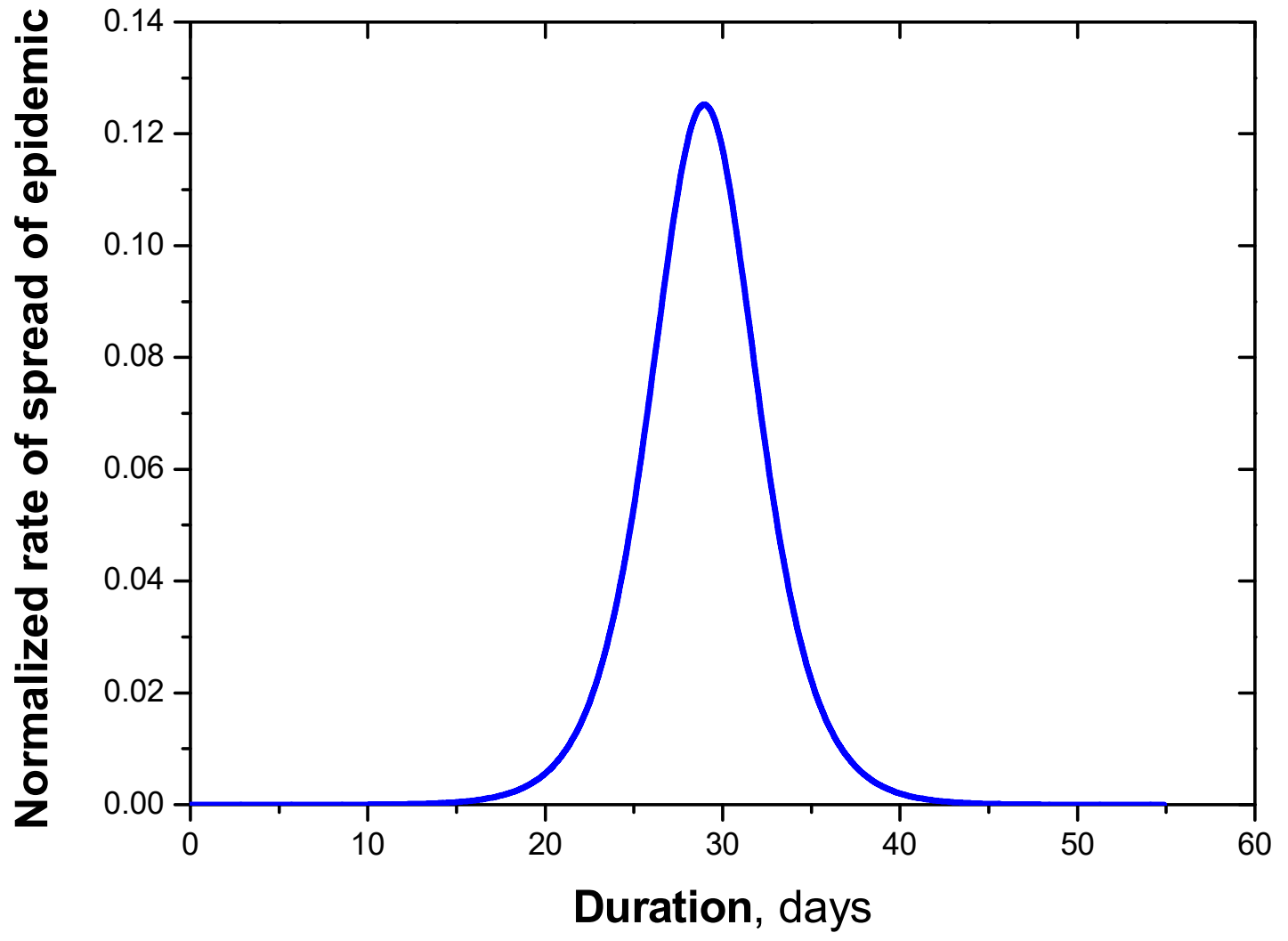

Figure 3. The rate of epidemic spread in the non-oscillating regime

For $kb_a = 0.5$ the averaged epidemic spread rate normalized to the population size $N$ is displayed in Fig.3. It is obvious from comparison of Fig. 2 and Fig. 3 that the maximum deviations of the number of sick people from the non-oscillating regime occur near the maximum of the averaged spreading rate. Indeed, near the maximum of the epidemic propagation rate, the right-hand side of equation (1) approaches zero, and even small fluctuations make a noticeable contribution to the process.

To process the primary statistical data on the number of sick people in practical work, the median smoothing procedure is used. For the parameter $kb_a$ = 0.5 the comparison of the number of cases of median smoothing (smoothing window size is taken 7 days and our approach with the selection of a sinusoidal term is shown. It can be seen that the median smoothing and our approach give practically the same result. Median smoothing was performed using the Mathcad 15 program. Nevertheless, the median smoothing with a window of 7 days retains a weak trace of fluctuations.

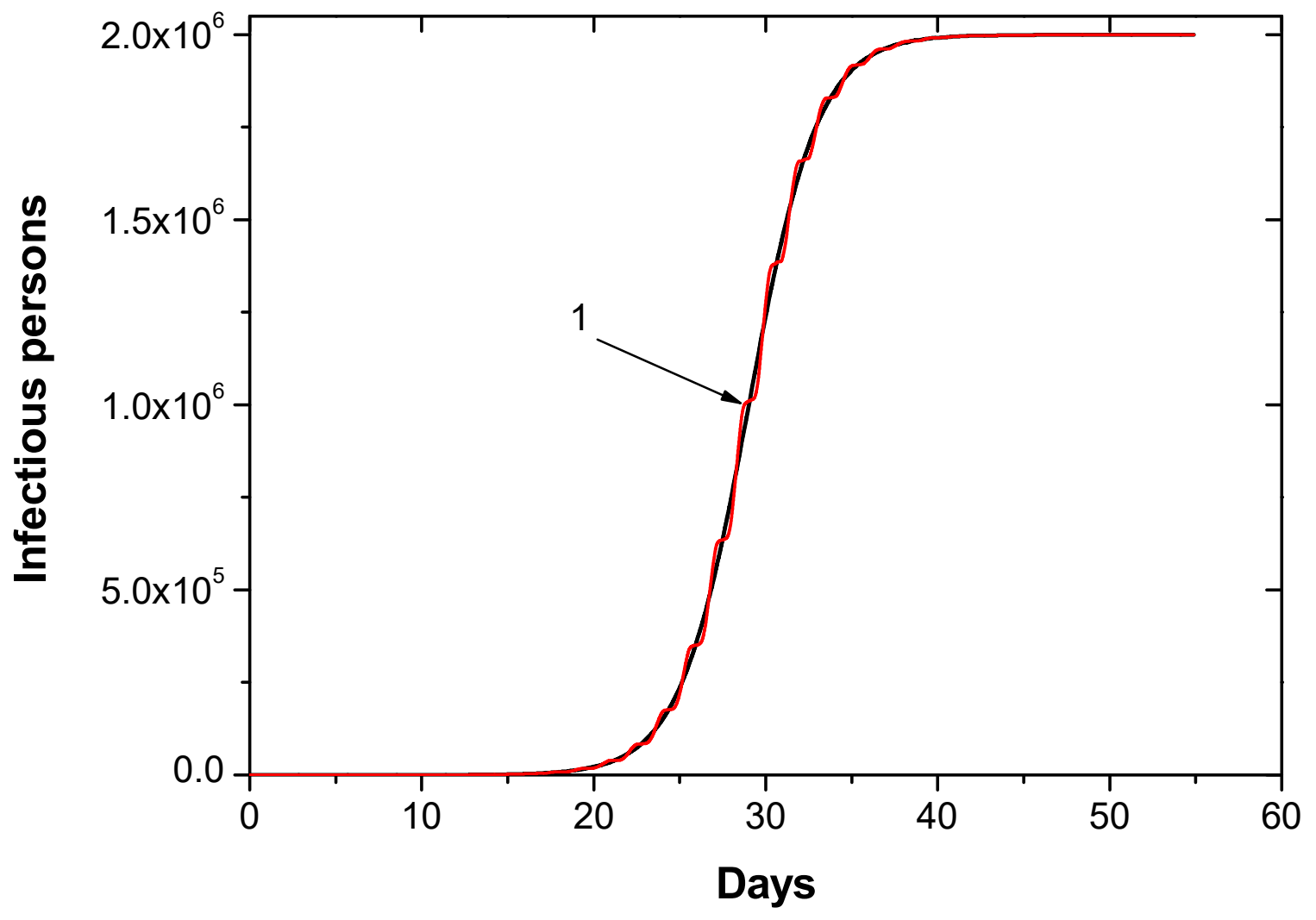

Fig. 4 Median smoothing

Curve 1 (red) is the median smoothing with the window width of 7 days, curve 2 is for $y_s$

When the smoothing window is increased to 14 days, the trace of fluctuations disappears.

## Conclusions

The number of daily contacts among people in any population is a stochastic variable which fluctuates around the mean value. Naturally, this phenomenon also manifests itself during the spread of epidemic. In particular, it causes fluctuations in the number of infected people. This effect is evident for the COVID-19 epidemic [1]. Fluctuations in the number of infected people affect the process of hospitalization of patients.

The mathematical model of the spread of the epidemic is developed, which considers fluctuations in the number of human contacts. The time-fluctuating variable of the number of social contacts $b$(t) was represented as the sum of the first two terms of the Fourier series expansion of number of contacts, expression (3). As the illustration for one population numerical solution is obtained. The parameter $b_a$ has the main influence on the duration of the epidemic.

It is shown that the largest fluctuation amplitudes occur when the right-hand side of the logistic equation is close to zero. In other words, near the maximum rate of the epidemic spread. After that, the effect of fluctuations decreases until the end of the epidemic.

It can be assumed that the increase in the amplitude of fluctuations in the number of infected people indicates that the epidemic is passing through the maximum rate of spread in the population.

To solve the inverse problem of determination of the parameters in equation (1) using primary medical information [17], it is recommended to use expression (3). Moreover, the fundamental frequency $\omega$ and phase $\varphi$ should be determined without using the procedure of primary data smoothing.

From a practical point of view it is shown that the median smoothing with the width of smoothing window of 7 days and our approach with two terms of the Fourier series expansion give practically identical results. The theoretical advantages of our approach are obvious.